\documentclass[prd,aps,showpacs,floats]{revtex4}
\usepackage{amsmath,amssymb}
\usepackage{graphicx}
\usepackage{bm}
\newcommand{\nova}{\mbox{\hbox{NO}$\nu$\hbox{A}}}

\newcommand{\myodot}{{\mathchoice
    {\raisebox{\depth}{$\displaystyle\odot$}}
    {\raisebox{\depth}{$\textstyle\odot$}}
    {\raisebox{\depth}{$\scriptscriptstyle\odot$}}
    {\raisebox{\depth}{$\scriptscriptstyle\odot$}}
    }}

\begin{document}

\title[Short Title]{Random magnetic fields inducing solar neutrino
spin-flavor precession in a three generation context}
\author{M. M. Guzzo$^1$}\email{guzzo@ifi.unicamp.br} 
\author{P. C. de Holanda$^{1,2}$}\email{holanda@ifi.unicamp.br}
\author{O.  L. G. Peres$^1$}\email{orlando@ifi.unicamp.br} 
\affiliation{
  $^1$ Instituto de F\'\i sica Gleb Wataghin - UNICAMP, 
  13083-970 Campinas SP, Brazil \\
  $^2$ Instituto de F\'\i sica, Universidade de S\~ao Paulo, 
  05315-970, S\~ao Paulo SP, Brazil 
} 
\newcommand{\lsim}{\,\lower .5ex\hbox{$\buildrel < \over {\sim}$}\,}
\newcommand{\gsim}{\,\lower .5ex\hbox{$\buildrel > \over {\sim}$}\,}
\newcommand{\vect}[1]{\overrightarrow{\sf #1}}
\newcommand{\system}[1]{\left\{\matrix{#1}\right.}
\newcommand{\displayfrac}[2]{\frac{\displaystyle #1}{\displaystyle #2}}
\newcommand{\nucl}[2]{{}^{#1}\mbox{#2}}
\newcommand{\diff}{{\rm\,d}}
\newcommand{\ea}{{\em et al.}}
\newcommand{\be}{\begin{equation}}
\newcommand{\ee}{\end{equation}}
\begin{abstract}
We study the effect of random magnetic fields in the spin-flavor
precession of solar neutrinos in a three generation context, when a
non-vanishing transition magnetic moment is assumed. 
While this kind of precession is strongly constrained when the
magnetic moment involves the first family, such constraints 
do not apply if we suppose a transition magnetic moment between the second and third 
families. In this scenario we can have a large non-electron anti-neutrino 
flux arriving on Earth, which can lead to some interesting
phenomenological consequences, as, for instance, the suppression of day-night
asymmetry. We have analyzed the high energy solar neutrino data and
the KamLAND experiment to constrain the solar mixing angle, 
$\tan \theta_{\myodot}$,  and solar mass difference, $\Delta
m_{\myodot}^2$, and we have found a larger shift of allowed values.
\end{abstract}

\maketitle

\baselineskip=20pt

\section{Introduction}

Recent results of KamLAND experiment~\cite{Kam} confirmed the Large
Mixing Angle (LMA) realization of the MSW phenomenon as the
explanation to the solar neutrino
anomaly~\cite{Cl,sage,gallex,gno,SK,SK2,sno,sno-all}.
Furthermore, KamLAND results rule out several other possible solutions
to the solar neutrino problem based on exotic phenomena~\cite{Gago:2001si},
like as the resonant spin-flip
conversion~\cite{Lim:1987tk,Akhmedov:1988uk,Balantekin:1990jg,Guzzo:1998sb,Gago:2001si,Akhmedov:2002mf,Friedland:2002pg,pulido2004}
induced by a non-vanishing neutrino magnetic moment interacting 
with solar magnetic fields.

Nevertheless exotic phenomena can generate sub-leading effects which
are still allowed by present solar neutrino data. Such effects can add
new features to this picture, in particular, changing the
determination of the neutrino oscillation parameters. Examples of
these sub-leading effects were analyzed in~\cite{Guzzo:2003xk}, where
random fluctuations of solar matter were considered, or
in~\cite{mswnsni,Guzzo:2004ue}, where non-standard neutrino
interactions induced a different determination of the oscillation
parameters necessary for a solution to the solar neutrino anomaly.
Here we study another possible sub-leading effect: the consequences
of neutrino interaction with {\it random} solar magnetic fields
through a non-vanishing magnetic moment.

The random magnetic scenario was studied in a different 
context~\cite{Pastor:1995vn,Semikoz:1998ef,Sahu:1998jh,Bykov:1998gv,enqvist-semikoz,Miranda:2003yh,Tortola:2004vh}
but always assuming a magnetic moment linking the electron neutrino
with the muon and tau anti-neutrino families. In this framework 
electron anti-neutrinos are produced as a consequence of the spin-flavor
precession, due the large mixing angles $\theta_{\odot}$ and
$\theta_{atm}$, the first one coming from solar neutrino analysis and
the second one from atmospheric neutrino data. Since the solar
electron anti-neutrino flux is strongly constrained by
data~\cite{Gando:2002ub,Eguchi:2003gg,Miranda:2003yh,Liu:2004ny,Tortola:2004vh},
that analysis puts severe limits on the size of the magnetic moment
(assuming a particular solar magnetic field profile), in order to
avoid strong spin-flavor conversion producing a sizable anti-electron
neutrino flux. Recently also a limit for anti-non-electron neutrinos
was quoted~\cite{pulido2004}, 
but these limits are weak and does not impose any constrain in our analysis.

For the neutrino parameters in MSW-LMA region, 
the spin-flavor precession is very small  
for typical values of magnetic field in the Sun. However, it was recently
pointed out~\cite{Miranda:2003yh,Tortola:2004vh} that
random magnetic field could enhance this conversion. Consequently, stronger limits for  the neutrino magnetic momentum $\mu$ was obtained, typically,  $0.78-1.2 \times 10^{-10} \mu_{B}$~\cite{Tortola:2004vh}

A conveniently chosen non-vanishing magnetic moment in the muon-tau sector
leads to a very different scenario. Tau anti-neutrinos are produced
through $\nu_\mu\rightarrow\bar{\nu}_\tau$ conversion, and
assuming a vanishing mixing angle $\theta_{13}$, the production of
electron anti-neutrino is
kept very small. 
The final solar neutrino flux can be a mixing of 
$\nu_e$, $\nu_\mu$, $\nu_\tau$,  $\bar{\nu}_\mu$ and $\bar{\nu}_\tau$, which 
can have some interesting phenomenological consequences. 
The more direct one would be a correlation between the solar magnetic field 
and the proportion between the different neutrino families.
Also the regeneration effect will be modified due to a different proportion 
of active neutrinos in the solar mass eigenstates, in analogy with the effect 
of a non-vanishing $\theta_{13}$~\cite{Akhmedov:2004}.  The next round of
reactor and long-baseline experiments can measure the $\theta_{13}$ angle if
$\sin^2(2\theta_{13})>0.01$.

We analyze here the scenario where neutrinos interact with {\em random} solar magnetic fields trough  a
non-vanishing magnetic moment between  anti-muon and anti-tau-neutrinos as a sub-leading effect in the context of LMA solution to the solar neutrino anomaly. We 
combine the results of this analysis with the constrains coming from the KamLAND observations.

\section{Formalism} \label{formalism}

We start working in a 6$\times$6 matrix formalism, with 
$\nu=(\nu_e,\nu_\mu,\nu_\tau,\bar{\nu}_e,\bar{\nu}_\mu,\bar{\nu}_\tau)^T$ where 
we include, besides the 
usual mass induced oscillation, magnetic moment terms between second and 
third families. We use as the mixing matrix the standard PMNS
(Pontecorvo, Maki, Nakata,Sakata)  mixing
matrix as presented in the Particle Data Group (PDG)~\cite{pdg}. 
After rotating out the angle 
$\theta_{atm}\equiv\theta_{23}$, we can decouple the first, second and sixth families,
obtaining: 
\begin{eqnarray}
& 
i{\frac{\displaystyle{\ d }}{\displaystyle{\ dt}}}
\left(
\begin{array}{c}
\nu_e \\ \nu_\mu' \\ \bar{\nu}'_\tau
\end{array}
\right) = 
& 
\left(
\begin{array}{ccc}
-\delta c_{2\theta} +V_e+V_\mu & \delta s_{2\theta}          &  0     
           \\

\delta s_{2\theta}             & \delta c_{2\theta} + V_\mu  & \mu B\exp(i\alpha)\\
 0                          & \mu B\exp(-i\alpha)      & \Delta - V_\mu
\end{array}
\right) \left(
\begin{array}{c}
\nu_e \\ \nu_\mu' \\ \bar{\nu}'_\tau
\end{array}
\right),
\label{motion}
\end{eqnarray}
where $\delta=\frac{\Delta m^2_{21}}{4E}$, 
$\Delta=\frac{\Delta m^2_{32}+\Delta m^2_{31}}{4E}$, $\Delta m^2_{ij}$ 
is the mass squared difference between neutrino families $i$ and $j$, 
$V_e$ and $V_\mu$
are the matter potentials, $\alpha$ is a phase of magnetic field,
$c_{2\theta}$ and $s_{2\theta}$ are cosine and sine of solar angle
$\theta_{\odot}$. The eigenstates 
$\nu_\mu'$ and  $\bar{\nu}'_\tau$ are linear combinations of weak
states as $\nu_\mu'=c_{\theta_{23}}\nu_\mu +s_{\theta_{23}}\nu_\tau$, 
$\nu_\tau'=-s_{\theta_{23}}\nu_\mu +c_{\theta_{23}}\nu_\tau$. (From now
on we suppress the prime symbol.)

It is more convenient to work in matrix density formalism, where the 
effects of random magnetic fields can be included in the evolution equation. 
The evolution equation, as given in Eq.~(\ref{motion}), can be rewritten
using the formalism of the density matrix $\rho$, with elements
$\rho_{ij} \equiv |\nu_i><\nu_j|,i,j=1,..,3$. 
We can expand the resulting
$3\times 3$ matrix using a complete set of
$3\times 3$ matrices, : $\lambda_{\nu}$, $(\nu=0,...,8)$: the $\lambda_0=\sqrt{2/3} I_3$
( $I_3$ is the $3\times3$ identity matrix) and $\lambda_{\nu}(\nu=1,...,8)$ are the 
Gell-Mann matrices. We assume Tr$(\lambda_{\nu}
\lambda_{\mu}) =2 \delta_{\nu\mu}$. The final equation can be written as
\begin{eqnarray}
&&\frac{\partial \rho_\mu}{\partial t} = \sum_{\nu\alpha}
h_{\nu}\rho_{\alpha} {f}_{\nu\alpha\mu}  
+ \sum_{\nu} {\cal L}_{\mu\nu}\rho_\nu~, \quad \mu, \nu , \alpha = 0, \dots 8 \nonumber \\
\label{expandedlind}
\end{eqnarray}
where the $h_{\nu}\equiv$ Trace$(H \lambda_{\nu})/2$ are defined as
$H=\sum_{{\nu}=0,8}h_{\nu}\lambda_{\nu}$, where  the elements
$h_2$,$h_4$ and  $h_5$  vanish. Similarly, $ \rho = \sum_{\nu}\rho_{\nu} \lambda_{\nu} $. 
Explicitly the coefficients $h_i$ are 
\begin{eqnarray}
h_0 =  \frac{\Delta}{3} + \frac{V_e+V_\mu}{3}, 
                \quad\quad && h_6  =+ \mu B \cos\alpha, \nonumber \\
h_1 =\delta s_{2\theta},  
                \quad\quad && h_7  =- \mu B \sin\alpha, \nonumber \\
h_3 = -\delta c_{2\theta}+\frac{V_e}{2},  \quad\quad 
&& h_8 = -\frac{\Delta}{\sqrt{3}} + \frac{1}{2\sqrt{3}}(V_e+4V_\mu).
\end{eqnarray}

If ${\cal L}_{\mu\nu}$ is identically zero, we have a Liouville equation,
that after some algebra can be written as
\begin{eqnarray}
& 
{\frac{\displaystyle{\ d }}{\displaystyle{\ dt}}}
\left(
\begin{array}{c}
\rho_1 \\ \rho_2 \\ \rho_3 \\ \rho_4 \\ \rho_5\\ \rho_6\\ \rho_7\\ \rho_8 
\end{array}
\right) = 
& 
\left(
\begin{array}{cccccccc}
0    &-2h_3 &0     &-h_7            &h_6   &0        &0     &0          \\
2h_3 &0     &-2h_1 &-h_6            &-h_7  &0        &0     &0          \\
0    &2h_1  &0     &0               &0     &h_7      &-h_6  &0          \\
 h_7 &h_6   &0     &0    &-h_3-\sqrt{3}h_8 &0        &-h_1   &0         \\
-h_6 &h_7   &0     &h_3+\sqrt{3}h_8 &0     &h_1      &0       &0        \\
0    &0     &-h_7  &0       &-h_1  &0      &h_3-\sqrt{3}h_8 &\sqrt{3}h_7 \\
0    &0     & h_6  &h_1     &0     &-h_3+\sqrt{3}h_8 &0   &-\sqrt{3}h_6  \\
0    &0     &0     &0       &0     &-\sqrt{3}h_7     &\sqrt{3}h_6  &0    \\
\end{array}
\right) 
\left(
\begin{array}{c}
\rho_1 \\ \rho_2 \\ \rho_3 \\ \rho_4 \\ \rho_5\\ \rho_6\\ \rho_7\\ \rho_8 
\end{array}
\right) ,
\label{evolution}
\end{eqnarray}
where the matrix is
antisymmetric as it is expected in the Liouville equation. 

Until now we are only dealing with the usual MSW mechanism with spin-flip
terms via regular magnetic fields (see Eq.~(\ref{motion})).
Note that or $h_6 \propto h_7 \propto  \mu\rightarrow 0$ we recover the usual MSW LMA mechanism.

\subsection{Magnetic Field Profile}

In order to quantitatively perform the analysis, one has to choose a solar magnetic field profile.
We assume for the magnetic field a triangular profile in the convective zone, 
with a maximum at $r/R_{Sun}=0.85$ of $B_{MAX}=100$ kG, and zero in the 
radiative zone. We assume that the magnetic field will be composed by
a regular part and a random part. For values of oscillation parameters
in the LMA region, 
the contribution of the
regular field is completely irrelevant. However, the random
character of magnetic field allows for large amplitudes changes that
can significantly modify the neutrino evolution inside sun. 

We will introduce the random features in the magnetic field 
through a delta-correlated fluctuations~\cite{Guzzo:2003xk,Nunokawa:1996qu}, 
which has the advantage of allowing a simple analytical parametrization of the
random field in the neutrino evolution
equation.

The system evolution can than be divided in 
two parts: first the simple MSW conversion in the production region of 
solar neutrinos, where we can take the formulas for the two families 
conversion, as presented, for instance, in~\cite{deHolanda:2004fd}. For 
$r/R_{Sun}>0.7$ the magnetic field starts to act on the system and
then the conversion probabilities will depend of the neutrino magnetic
momentum $\mu$.

The random features will be introduced through the ${\cal L}_{\mu\nu}$
piece of neutrino evolution in the matrix density formalism,
Eq.~(\ref{expandedlind}). In more general formalism, we should
consider the relative size of coherent length of the magnetic field
that we call $L_0$ and the neutrino oscillation length,
$\lambda_\nu\equiv \pi\frac{4E}{\Delta m_{21}^2}$. 
The condition to have a decoupling between the LMA MSW oscillations
and the spin-flip induced by random magnetic fields is
defined as
\begin{equation}
\lambda_\nu >>L_0~~.
\label{condition}
\end{equation}

Following Ref~\cite{Nunokawa:1996qu}, 
after some algebra, the change in the neutrino evolution to the random
character can be written as 
follows:
\begin{eqnarray}
\rho_{11}=\rho_{22}=\rho_{33}=\rho_{44}=\rho_{55}=\rho_{66}/2=\rho_{88}/3=-2k
, \nonumber \\
\rho_{38}=\rho_{83}=2\sqrt{3}k~~,
\label{randomentries}
\end{eqnarray}
where:
\[
k=<(\mu \bar{B}_x)^2>L_0=<(\mu \bar{B}_y)^2>L_0~~, 
\]
where $\bar{B}_{x,y}$ are the random components of the magnetic field
perpendicular to neutrino trajetory. We take this components to be 
proportional to the regular magnetic field.

To analyze if the values used for the parameter $k$ are reasonable, we
can write it in convenient units:
\[
k=1.7 10^{-17}\left[\frac{\mu}{10^{-11}\mu_B}\right]^2 
\left[\frac{B}{1MG}\right]^2\left[\frac{L_0}{1km}\right]  {\rm eV}~~.
\]
If we take the neutrino oscillation parameters from the best fit point
of the standard solar neutrino analysis,
$(\tan^2\theta_{\myodot},\Delta m^2_{\myodot})=
(0.4,8\times 10^{-5}~{\rm eV^2})$, we have that the oscillation length
for a $10$ MeV neutrino is $\lambda_\nu\sim200$ km.

The probabilities can be  treated
classically since the averaging 
over the production region suppresses any interference effect. 
We calculate the final survival probability using:
\begin{eqnarray}
P_{ee} &=& P_{e1}^{rad}P_{1e}^{conv}+ P_{e2}^{rad}P_{2e}^{conv}+ 
           P_{e3}^{rad}P_{3e}^{conv}~~, \nonumber \\
P_{e\mu} &=& P_{e1}^{rad}P_{1\mu}^{conv}+ P_{e2}^{rad}P_{2\mu}^{conv}+ 
           P_{e3}^{rad}P_{3\mu}^{conv}~~, \nonumber \\
P_{e\tau} &=& P_{e1}^{rad}P_{1\tau}^{conv}+ P_{e2}^{rad}P_{2\tau}^{conv}+ 
           P_{e3}^{rad}P_{3\tau}^{conv}~~,
\end{eqnarray}
where $P_{e1}^{rad}$ is the probability that an electronic neutrino $\nu_e$ 
arrives at the bottom of the convective zone as $\nu_1$, $P_{1e}^{conv}$ is 
the probability that a $\nu_1$ crosses the convective zone and leave the sun 
as $\nu_e$. Since we have $B=0$ in the radiative zone, $P_{e3}^{rad}=0$.
The probabilities  $P_{e1}^{rad}$ and $P_{e2}^{rad}$ are equal to
computed using the $2\times 2$ evolution equation for the LMA MSW
mechanism.

\begin{figure*}
\includegraphics[height=12cm,width=15cm]{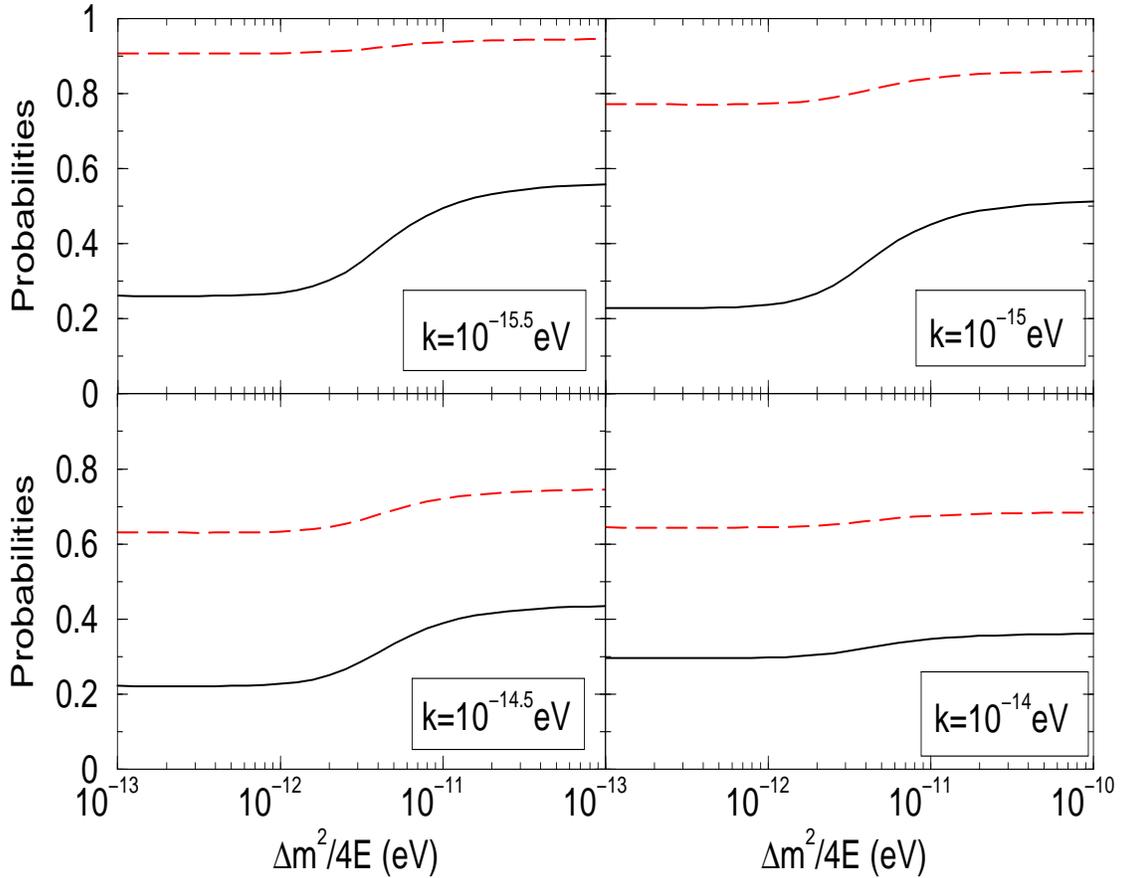} 
\caption{Neutrino survival probabilities, the solid line is $P_{ee}$,
the dashed is the sum $P_{ee}+P_{e\mu}$, and the $P_{e\bar{\tau}}$ is
the remaining until the maximum. We used $\tan^2\theta=0.4$ in all panels.}
\label{fig:probability}
\end{figure*}

Now we are in condition to compute the $P_{1e}^{conv}$ and
$P_{2e}^{conv}$ probabilities. In the convective region, the regular
magnetic field is too small to induce spin-flip conversion and then
the elements $h_6$ and $h_7$ vanishes. Then the Eq.~(\ref{evolution})
decouples in a sub-sector containing only
$(\rho_1,\rho_2,\rho_3,\rho_8)$ elements as follows:

\begin{eqnarray}
& 
{\frac{\displaystyle{\ d }}{\displaystyle{\ dt}}}
\left(
\begin{array}{c}
\rho_1 \\ \rho_2 \\ \rho_3 \\ \rho_8 
\end{array}
\right) = 
& 
\left(
\begin{array}{cccccccc}
-2k  &-2h_3 &0          &0           \\
2h_3 &-2k   &-2h_1      &0           \\
0    &2h_1  &-2k        &2\sqrt{3}k  \\
0    &0     &2\sqrt{3}k &-6k         \\
\end{array}
\right)
\left(
\begin{array}{c}
\rho_1 \\ \rho_2 \\ \rho_3 \\ \rho_8 
\end{array}
\right) .
\label{evol2}
\end{eqnarray}

It is interesting to note that the evolution equation does not depend
on $h_8$, not depending therefore on the atmospheric mass scale $\Delta
m_{32}^2$. This is an important feature of our configuration, since as
mentioned in the last section, the condition of validity of our
equation is that the coherence length is smaller then the neutrino
oscillation length.
\\

At the bottom of convective zone the initial conditions are 
\begin{eqnarray}
\nu_1&:& \rho_0(0) = 1/3,\,\, \rho_1(0) = -\sin{2\theta}/2,\,\, 
      \rho_3(0) =\cos{2\theta}/2, \,\,\rho_8(0) = 1/(2\sqrt{3}) \nonumber \\
\nu_2&:& \rho_0(0) = 1/3,\,\, \rho_1(0) = \sin{2\theta}/2,\,\, 
      \rho_3(0) =-\cos{2\theta}/2,\,\, \rho_8(0) = 1/(2\sqrt{3}) 
\end{eqnarray}

To study the anti-neutrino production, we solve numerically
Eq.~(\ref{evol2}) for different values of $k$, and with initial
conditions given by the usual MSW mechanism in the radiative zone. 

The effect of the random magnetic field inclusion in evolution equation 
can be seen in Fig.~\ref{fig:probability}, where we can read the fraction of 
the different neutrino flavors, 
$\nu_e:\nu_{\mu}:\bar{\nu}_{\tau}$. The solid line represents the electronic 
neutrino survival probability, $P_{ee}$,  while the dashed line is the
$\nu_e+\nu_\mu$ fraction, $P_{ee}+P_{e\mu}$ The remaining until the
no-oscillation value is the contribution of the $P_{e\bar{\tau}}$
probability. In all panels we used $\tan^2\theta=0.4$.

For small values of $k$, the electronic and muonic neutrinos start to 
convert into $\bar{\nu}_{\tau}$, and, as a consequence, the electronic 
survival probability decreases.
For large values of $k$, the neutrino flux tends to split equally in 
the three neutrinos flavors, with $~1/3$ of the total flux for each flavor. 

Taking the marginal allowed value of $L_0=10$ km, in order
to have $k=10^{-14}$ eV as in the last panel of
Fig.~\ref{fig:probability}, we should have $B\sim 10$ MG, which is
hardly acceptable. Lower values of neutrino energy will decrease the
neutrino oscillation length, making it more difficult to fulfill the
conditions in Eq.~(\ref{condition}). In this
sense, the values of $k=10^{-15}$ eV seems more feasible in a realistic
scenario. 

However, as pointed out in Section~\ref{formalism}, we expect that the
effects of neutrino conversion would be stronger when 
$\lambda_\nu\sim L_0$. So the limitations expressed in the last
paragraph are only numerical limitations, and not physical
constraints. Having this in mind, we decided to extend the analysis up to 
$k=10^{-14}$ eV, assuming that a complete numerical integration of 
Eqs.~(\ref{evolution}) and (\ref{randomentries}) would give qualitatively 
the same results as we present here.

\section{LMA region}

The validity of our numerical treatment is energy dependent, so we have 
made the choice to limit our fit to solar neutrino data to a specific 
neutrino energy range. Since the validity of our approximations may not 
hold for low energy neutrinos, we included only the data for the high energy 
neutrino (SNO-I, SNO-II and SK). We must be careful in analyzing
allowed regions for neutrino parameters, since the inclusion of low
energy solar neutrino data should change this picture.

\begin{figure*}

\includegraphics[height=12cm,width=15cm]{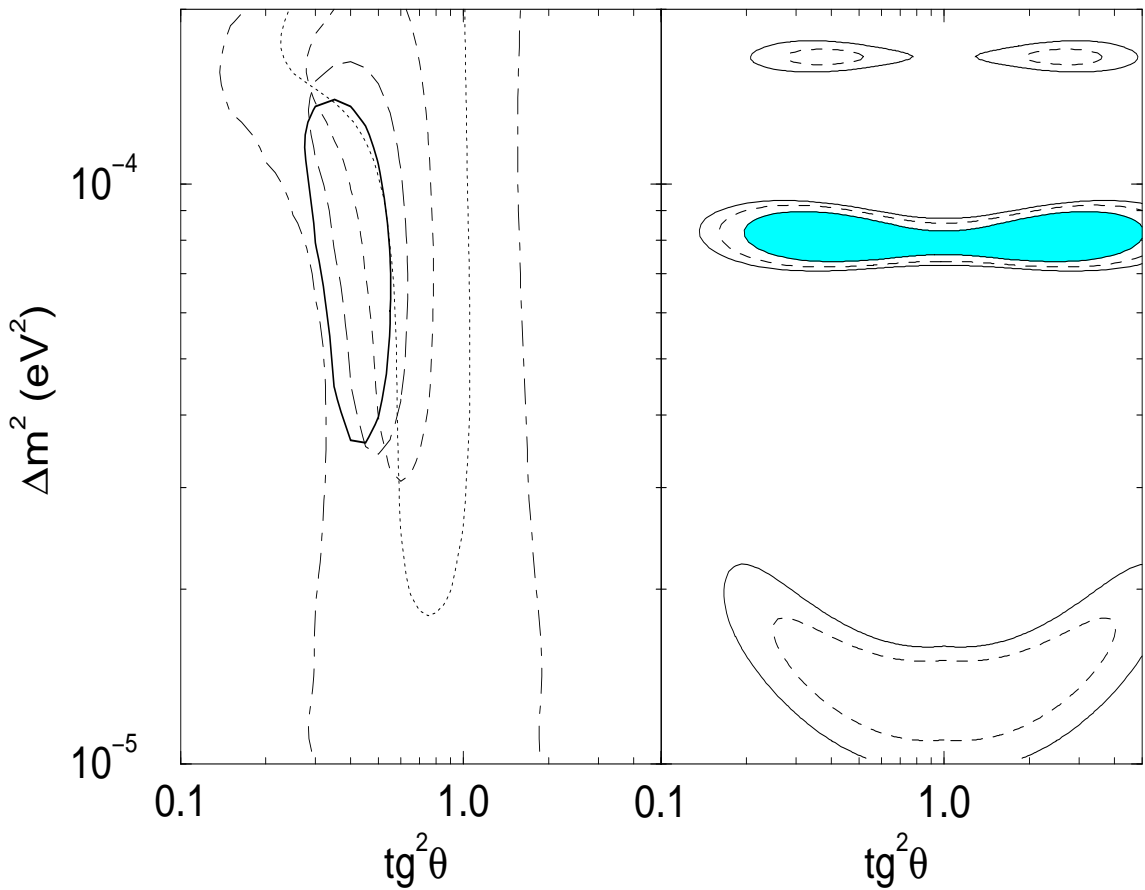}, 
\caption{The LMA compatibility region for SNO+SK data (left panel) 
and the KamLAND allowed region (right panel). In left panel, the black line 
stands for no magnetic field, the long-dashed line for $k=10^{-15.5}$,
the short-dashed line for $k=10^{-15}$, the dotted line for
$k=10^{-14.5}$ and the dot-dashed line for $k=10^{-14}$.
In the right panel the allowed regions stands for 95\% C.L., 99\% C.L. and 
3$\sigma$ for respectively the filled, dashed and black curves.}
\label{fig:lmaregion}
\end{figure*}

The effect of the random magnetic field in the LMA region can be seen in 
Fig.~\ref{fig:lmaregion}. As the parameter $k$ increases, the
electronic survival probability decreases. This effect can be
compensated by a higher value of neutrino mixing angle, moving the
allowed region to the right in left panel of
Fig.~\ref{fig:lmaregion}.  

For larger values of $k$, all probabilities tend to $~1/3$, with a
weak dependence of the mixing angle. Since the Super-Kamiokande and
SNO 
results are in accordance with this probability, in this scenario even
maximal mixing is allowed. But also a interesting phenomenon
occurs. Now a significant part of the total neutrino flux does not
take part in regeneration effect in Earth. Actually, if we have
exactly a equally equipartition of $\nu_e$, $\nu_\mu$ and
$\bar\nu_\tau$ fluxes the regeneration effect vanishes, regardless of
the neutrino mass difference $\Delta m^2$.

The right panel of Fig.~\ref{fig:lmaregion} presents the KamLAND
allowed regions, for $95\%$ C.L., $99\%$ C.L. and $3\sigma$. 
Maximal mixing is allowed at $62.1\%$ C.L., and low values of $\Delta
m^2$ are still consistent with data at $99\%$ C.L.. This last region
is inconsistent with solar neutrino experiments because it
predicts a too strong regeneration effect, not seen by data.

In the present context, the regeneration effect can be suppressed for
large values of $k$. As a result, lower values of $\Delta m^2$ are
allowed, and a new region of compatibility between solar and KamLAND
data appears. In other contexts of non-standard neutrino
physics~\cite{Guzzo:2003xk,Guzzo:2004ue}, we have called this region
very-low LMA. 

\begin{figure*}
\includegraphics[height=12cm,width=15cm]{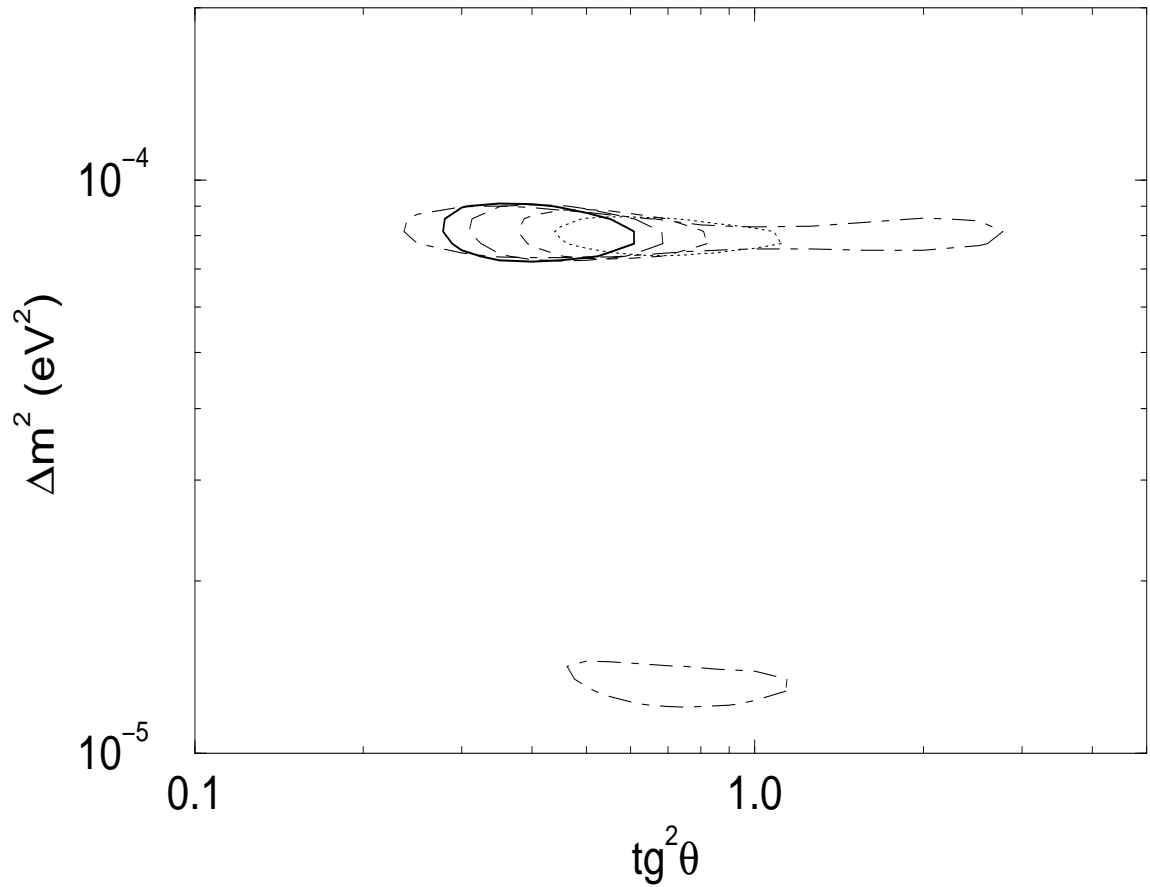}, 
\caption{Combined result for SNO+SK and KamLAND data, at 99\% C.L., for the 
same values of $k$ presented in Fig.~\ref{fig:lmaregion}. Maximal mixing is 
allowed for $k>10^{-14.5}$.}
\label{fig:lmakl}
\end{figure*}

In Fig.~\ref{fig:lmakl} we present the allowed region for a combined
analysis of high-energy solar neutrino and KamLAND data. 
We can see in this plot both the displacement
of the allowed region to higher values of $\tan^2\theta$ for moderate
values of $k$, and the appearance of 
the very-low LMA region at $\Delta m^2\sim [1-2]\times10^{-5}$ eV$^2$
for $k=10^{-14}$ eV.

\section{$\theta_{13}\ne 0$}

In this work, we assumed $\theta_{13}= 0$, since non-vanishing 
values of this angle will lead to a production of $\bar{\nu}_e$, 
which is strongly constrained by data. Then, if $\theta_{13}\ne 0$ a limit in $\mu_{23}$ could be 
established in the same way the limits of the other components of magnetic 
moments were found~\cite{Miranda:2003yh}, in a two families analysis,  
denoted here by $\mu_\nu^{2fam}$. 

Including the angle $\theta_{13}$ in the mixing matrix would lead to 
the term $\mu_\nu\sin\theta_{13}$ connecting $\bar{\nu}_e$ to the active 
neutrino families. So the limits in $\mu_\nu^{2fam}$  
could be scaled to a limit in $\mu_{23}$ if a positive measurement 
of $\theta_{13}$ is achieved in the next round of reactor~\cite{reactor13} 
and accelerator~\cite{accelerator} experiments. This limit would be weaker 
then the present limits in $\mu_\nu^{2fam}$ by a factor $1/\sin\theta_{13}$.

\section{Conclusions}
 
We have investigated new effects in solar neutrino phenomenology due
to interactions of these particles with a random solar magnetic field.
Considering the context of LMA realization of the MSW solution to the
solar neutrino anomaly we have analysed the neutrino spin-flavor
conversion phenomenon which appears as a subleading effect when a
non-vanishing neutrino magnetic moment linking the second and the
third families is assumed. Such a magnetic moment induces a large
conversion of solar neutrinos into non-electron anti-neutrino flux,
which is not severely constrained by the solar neutrino
observations. 

Since the mixing angle $\theta_{13}$ can be considered very small, 
this conversion is not followed by a production of electron
anti-neutrinos flux which, in contrast to non-electron anti-neutrinos,
is very constrained by data.
 
The results of our analyses of SNO+SK compatibility region 
indicate that in the presence of solar random magnetic fields 
the allowed region for $\Delta m^2$ becomes larger while higher 
values of $\theta_{12}$ are found. 

This is a consequence of the fact that, 
in a three neutrino family context, an electron-neutrino 
survival probability of $P~\sim 1/3$ is possible, even for 
$\tan^2\theta_{12}\sim 1$. In fact, when the random components 
of the solar magnetic field is large enough in such way that 
$k=10^{-14}$, values around $\Delta m^2\approx 10^{-5}$eV$^2$ 
are included in the region allowed by solar neutrino observations. 
Furtermore, different proportion of neutrinos and anti-neutrinos
suppresses the regeneration effect of the solar neutrinos crossing 
the Earth matter. As a consequence, a totally new region of 
compatibility between solar neutrinos and KamLAND, which we call 
very-low LMA, appears at 99\%~C.L. for small values of 
$\Delta m^2_{21}\sim [1-2]\times 10^{-5}$ eV$^2$ and maximal mixing.

\begin{acknowledgments} 
This work was partially supported by Funda\c{c}\~ao de Amparo
 \`a Pesquisa do Estado de S\~ao Paulo (FAPESP) and  Conselho
Nacional de Desenvolvimento Cient\'\i fico e Tecnol\'ogico (CNPq).
\end{acknowledgments}

\end{document}